\documentclass[journal]{IEEEtran}
\ifCLASSINFOpdf
\else
\fi

\usepackage{xcolor}
\usepackage{amsmath}
\usepackage{amsthm}
\usepackage{amssymb}
\usepackage{array}
\usepackage{url}
\usepackage{balance, acronym}
\usepackage{graphicx}
\usepackage{flushend}
\usepackage{subcaption} 
\begin{document}
\title{Communication, Computing, Caching, and Sensing for Next Generation Aerial Delivery Networks}

\author{Gunes~Karabulut~Kurt,~\IEEEmembership{Senior Member,~IEEE,}
          and~Halim~Yanikomeroglu,~\IEEEmembership{Fellow,~IEEE}% <-this % stops a space

 \thanks{G. Karabulut Kurt is with the \textcolor{black}{Poly-Grames Research Center, Department of Electrical Engineering,  Polytechnique Montr\'eal, Montr\'eal, Canada, e-mail: gunes.kurt@polymtl.ca}}
\thanks{H. Yanikomeroglu is with the Department of Systems and Computer Engineering, Carleton University, Ottawa, Canada, e-mail: halim@sce.carleton.ca}}


\markboth{Journal of \LaTeX\ Class Files,~Vol.~14, No.~8, August~2015}%
{Shell \MakeLowercase{\textit{et al.}}: Bare Demo of IEEEtran.cls for IEEE Journals}



\maketitle

\begin{abstract}
This paper describes the envisioned interactions between the information and communication technology  and  aerospace industries  to serve autonomous devices for next generation aerial parcel delivery networks. 
The autonomous features of  fleet elements of the delivery network are enabled by \textcolor{black}{the} increased throughput,  improved coverage, and  near-user computation capabilities of  vertical heterogeneous networks (VHetNets). A high altitude platform station (HAPS), located around 20~km above \textcolor{black}{the} ground level in a quasi-stationary manner, serves as the main enabler  of the vision we present. In addition to the sensing potential of the HAPS nodes, the use of communication, computing, and caching capabilities demonstrate the attainability of the ambitious goal of serving a fully autonomous aerial fleet capable of addressing instantaneous user demands and enabling supply chain management interactions with  delivery services in  low-latency settings. 
\end{abstract}

\begin{IEEEkeywords}
Aerial platooning, cargo drones, high altitude platform station (HAPS), vertical heterogeneous networks (VHetNets).
\end{IEEEkeywords}

% \fi
\IEEEpeerreviewmaketitle

\section{Introduction}
As e-commerce has emerged as an indispensable part of the retail sector, timely and efficient last-mile delivery solutions have materialized as its catalyst. A steady increase in the amount of e-commerce activities has been noted in numerous reports. An analysis by ACI Worldwide states that the transaction volumes in most retail sectors saw a 74\textcolor{black}{$\%$} increase in March 2020,  compared to the same period in 2019, due in part to the COVID-19 pandemic \cite{ACI}. The increase in e-commerce activities also increased the load on the delivery sector. The need for viable solutions to this problem became more apparent with skyrocketing demands and prolonged delivery times. To maintain  efficient operations, delivery services involving autonomous devices have emerged as a promising solution. In this view, conventional delivery services, involving cars \textcolor{black}{or} trucks, will be supplemented by emerging aerial delivery fleets. The autonomy of these devices will range  from a fully human operated level 0 to a fully autonomous level 5. 

The future of  delivery networks are expected to function based on the basis of combined operations between transportation and information and communication technology (ICT) networks. This will be supported by autonomous aerial delivery fleets operating over \textcolor{black}{the} designated airspace. We envision the joint use of cargo drones alongside hybrid or electric vertical take-off and landing (VTOL) aircraft, which can be used  both for the delivery of  parcels \textcolor{black}{and/or} the transportation of  cargo drones. 

Drone-supported aerial delivery networks have long been studied by  retail sector giants, including Amazon and Alibaba \cite{Nesrine}. However,  extensive use of such networks in metropolitan areas has not yet been considered. In trials by Amazon Prime Air and Tesco (in the UK), the target areas for drone-supported aerial deliveries are rural areas, where the settlements are sparse. Yet, the impact of such deliveries will be effective only by servicing the densely populated metropolitan areas. 

\begin{figure}[tb!]
    \centering
    \includegraphics[width=0.98\linewidth]{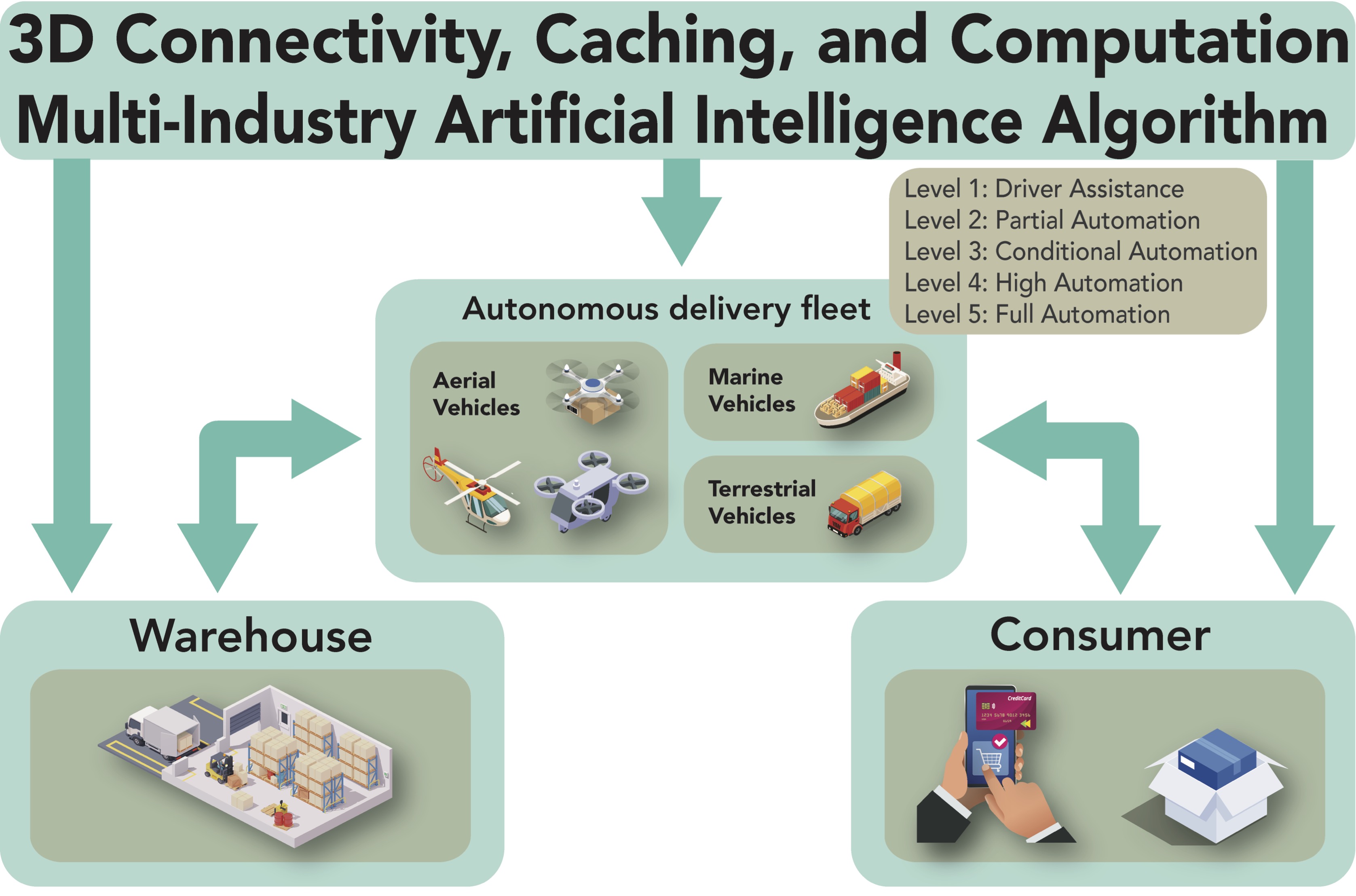}
\caption{An overview of the 3D connected smart delivery network. Real-time action is possible within the vehicles of the autonomous delivery fleet through the connected multi-industry artificial intelligence engine. A fully automated architecture is envisioned through the use of the autonomous delivery fleet.} \label{fig-ML}
\end{figure}

\textcolor{black}{The current } wireless network architecture \textcolor{black}{will} fail to provide sufficient support for for aerial vehicle assisted delivery solutions \textcolor{black}{due to the 3D natıre of the network}. Although high data rates are achievable for users on the ground and in highrises through meticulous planning of extant 5G networks, the coverage of up to \textcolor{black}{a} maximum permitted height  for  drone delivery nodes does not support high data rate and low-latency solutions \textcolor{black}{because of the non-isotropic radiation patterns of  terrestrial base station antennas}. The areas above the antennas do not receive coverage. It is expected that this lack of coverage will hinder the fully autonomous operations of drone delivery nodes. Furthermore, the next generation delivery networks are expected to be fully managed by artificial intelligence (AI) involving innovative distributed machine learning (ML) algorithms, as depicted in Fig.~\ref{fig-ML}. Yet the computational loads of these algorithms may be too intensive for  drone nodes, where energy efficiency will remain  a strict design goal\textcolor{black}{, hence these algorithms need to be executed at a diverse location}. 

To address these challenges, this paper examines offloading possibilities and delivery route planning for 3D highways within a vertical heterogeneous network (VHetNet) paradigm.  We present a realistic vision of a next-generation delivery network with a focus on issues pertaining to connectivity as well as computation\textcolor{black}{,}  caching\textcolor{black}{, and sensing}. \textcolor{black}{Among others, instantaneous navigation, traffic status, the energy level monitoring and the related prediction applications can be guided by the ML algorithms executed at the serving HAPS node. Furthermore in addition to location sensing, collection and gathering of the sensed data is possible at the HAPS node in terms of the sensing capabilities. Below, we describe the} main components of providing a fully connected, high rate, and low-latency 3D network. The solution we envision makes \textcolor{black}{the} use of high altitude platform station (HAPS) systems as an essential component. This architecture is also in line with the emerging literature on  6G networks. In our view, the main catalyzer will be HAPS constellations, which offer an excellent synergy between  evolving terrestrial networks and  emerging low Earth orbit (LEO) satellite constellations. By using HAPS for  connectivity, caching, and computational offloading, \textcolor{black}{we will have the opportunity to the enable all these functionalities at the network edge, hence we can significantly benefit from the expected performance improvement as clearly highlighted in \cite{e1} and quantified in \cite{e2}. With this feature,} we predict that next-generation delivery networks will soon become a reality, even in densely populated metropolitan areas. 

The rest of this paper is organized as follows. The evolution of  wireless architecture to support fully autonomous air fleets is described in Section II. \textcolor{black}{The} main architectural components are described from the perspectives of connectivity, caching,  computation\textcolor{black}{, and sensing}. Section III describes the main features of HAPS systems. In Section IV, we present our vision of AI-powered and connected delivery networks  with rapid response rates.  Open issues are highlighted in Section V. Section VI concludes the paper.  

\section{Towards 6G: The Interaction Between the ICT Network and the Delivery Network}

We envision that delivery networks in the near future will be semi-autonomous and that the goal of achieving full autonomy will be realized in the next two decades. A significant change in delivery networks has been the introduction of drone-based deliveries\textcolor{black}{. Numerous} trials mainly concentrate on rural areas with sparse populations and  housing \textcolor{black}{, however to be} economically viable,  next-generation last-mile delivery services need to address  metropolitan areas. Yet, the current ICT networks’ capabilities fail to address the operational needs of such delivery networks \textcolor{black}{due to the ambient} interference, shadowing, and a lack of global navigation satellite system (GNSS) signaling in urban corridors\textcolor{black}{, as noted}  in the trials of Unmanned Aircraft System Traffic Management (UTM), supported by the collaboration of NASA and FAA \cite{UTM}.  

The VHetNet paradigm, \textcolor{black}{has} the potential to address the needs of  next generation delivery networks. A VHetNet is composed of three layers: a terrestrial network, a space network (satellites), and an aerial network \cite{RAPOR1}. The terrestrial network is the main functional block of the VHetNet, which mainly connects users and devices to the core network. As the lowest layer, the terrestrial network includes various network generations, including 4G and 5G \textcolor{black}{networks}, in combination with unlicensed band systems, such as WiFi. Satellite networks are composed of three satellite layers\textcolor{black}{;} LEO, medium Earth orbit (MEO), and geosynchronous Earth orbit (GEO) satellite systems and their corresponding ground stations. %The interactions between these sub-layers from the ICT network's perspective is not yet applicable. GEO satellites are positioned at 35786 km above earth and they remain stationary with respect to ground. LEO and MEO satellites are positioned between 400-2000 km and 80000-20000 km above the ground. 
There are several commercially operated systems, including GEO and MEO constellations, which  mostly provide communication and surveillance/monitoring services. Forthcoming constellation deployment plans, including those of OneWeb, Amazon's Project Kuiper, and SpaceX, will introduce densely populated LEO constellations\textcolor{black}{.} %Hence, the interaction between these LEO constellations and terrestrial networks is expected  to increase in the very near future.

The aerial ICT network's architecture will consist of unmanned aerial vehicles (UAVs) in addition to airships, balloons, and HAPS systems. The drones are envisioned to be at a height of up to a few hundred meters. As for HAPS systems, the International Telecommunications Union (ITU) defines their operating altitude to be between 20 km and 50 km. However, most commercial HAPS trials target 18 km to 21 km, including Airbus Zephyr \textcolor{black}{and} Stratobus of ThalesGroup \cite{Survey}. The aerial network, which is connected to the terrestrial network, improves the flexibility of the network design in terms of both capacity and coverage. With an aerial ICT network, coverage of highly populated metropolitan areas will then be possible while supporting high data rates. Coverage of highly populated metropolitan areas \textcolor{black}{at high data rates}  will then be possible.

The aerial network in the VHetNet architecture needs to be carefully designed. \textcolor{black}{Two} interacting sub-layers in the aerial network will introduce  agility to the network functionalities. The first sub-layer includes the ultra-mobile UAV nodes, which can work as a base station, a relay node or a user equipment\textcolor{black}{, which can be considered a fleet element from our perspective}. The second complementary sub-layer is composed of HAPS systems,  the quasi-stationary network elements. \textcolor{black}{The relatively slow motion of a HAPS node can enable its connectivity to the core network  either through a radio frequency (RF) link or a free space optical (FSO) link.} This HAPS sub-layer will provide important functions in terms of coverage, computation,  caching, \textcolor{black}{and sensing. A HAPS node} has the potential to solve important problems in next-generation delivery networks, as detailed below.
 
In a VHetNet, the network elements mainly target three complementary objectives, addressing everything needed to make the 3D connected smart delivery network a reality:
\begin{enumerate} 
\item \textbf{Increased overall throughput:} The individual data rates and/or the total number of fleet elements that can be served can be increased by deploying new aerial base stations.  
\item \textbf{Improved coverage:} The outage probability can be reduced by the use of mobile base stations; hence,  coverage can be always provided to address the fully connected 3D networks to support autonomous delivery fleet elements. 
\item \textbf{Perform near-user computation:} The delay due to computation through the core network can be significantly reduced by performing the computation and caching functionalities near the fleet elements\textcolor{black}{, from the edge computing perspective}.
\end{enumerate} 

The management of this highly complex multi-connectivity network with multi-layer computation offloading is supported through the use of  ML  algorithms. Additionally, in a highly dynamic environment, classical radio resource management approaches may fail to address the tight quality of service (QoS) requirements of the \textcolor{black}{fleet elements}. Data-driven ML algorithms will serve as a solution to such problems, especially in a distributed sense \textcolor{black}{for each fleet element}, to address the quick decision making need of the next generation delivery networks. The three objectives listed above will be enabled by the VHetNet, and the HAPS components will serve as  indispensable element\textcolor{black}{s}.

\begin{figure}[tb!]
    \centering
    \includegraphics[width=0.98\linewidth]{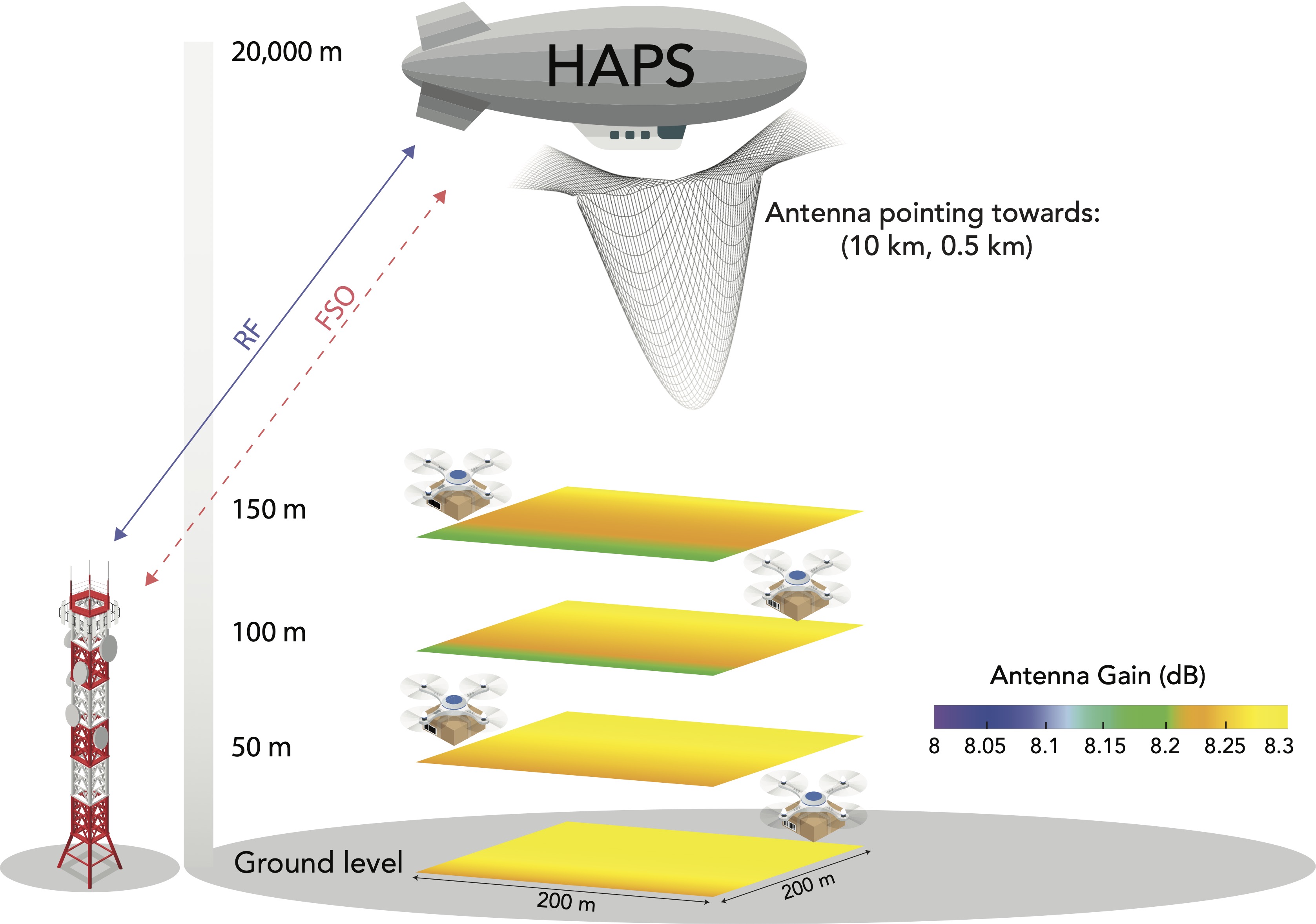}
\caption{A depiction of the antenna gain patterns of a HAPS node that is not beam-aligned.  The almost uniform gain observed due to the high-altitude of the HAPS mode provides a connectivity advantage with respect  to terrestrial base station towers and high speed LEO satellites. Radio frequency (RF) or free space optical (FSO) connections can be used for the connection to the terrestrial base station. The terrestrial base station interconnects the access network with the core network.}  \label{fig-HAPS1}
\end{figure}

\section{The Key Enabler: High Altitude Platform Station (HAPS)}  

Current ICT networks aim to provide coverage on the ground level, and inside buildings \cite{Mozaffari}. Such networks can not address the challenging requirements of  next generation delivery networks \textcolor{black}{according to the QoS requirements of each fleet element that can move in a 3D pattern}. One of the significant benefits of VHetNets is the quasi-stationary HAPS sub-layer of the aerial network, which may serve as an essential component for  aerial network planning and management. This can address the needs of next-generation delivery networks. An extensive description of the use of HAPS in VHetNets is given in \cite{Survey}.

Drones can travel up to a height of 121 meters (400 feet), \textcolor{black}{yet} there may not be coverage at this \textcolor{black}{height}. The reason is basic geometry: the antenna patterns are sectorized and directional, which means they do not transmit signals skyward. Although under ideal conditions 3D spherical coverage can be modeled, in practice no antenna can satisfy this ideal design. 3GPP study items TR 22.926 \textit{Guidelines for Extra-territorial 5G Systems} and TR 22.839 \textit{Study on vehicle-mounted relays} investigate  the use of  drone-mounted base stations\textcolor{black}{.} Although these \textcolor{black}{can} provide coverage for a specific height\textcolor{black}{,} the \textcolor{black}{contiguous} coverage probability at all heights is not feasible for  metropolitan areas.  The problem is mainly due to the beam patterns of base station antennas, which results in a high number of sidelobes\textcolor{black}{ in a 3D pattern.} This set-up also introduces another problem in the case of densely deployed devices with ultra-high mobility, such as swarms of drones, where the handoff of each node may introduce a substantial delay to the system. Considering  LEO satellites, although they can provide service to the operating altitudes of drones, the high speed of the satellites (with speeds of approximately 7 km/s) introduce a high load from a mobility management perspective. This leaves HAPS, with their quasi-stationary nature, as an indispensable enabler to address the requirements of  next generation delivery networks as opposed to the patchy coverage currently provided by  terrestrial \textcolor{black}{networks} \cite{Grace}. 

The signal transmission advantage, provided by the use of a HAPS as a mega-tower in the sky is depicted in Fig.~\ref{fig-HAPS1}. \textcolor{black}{The} antenna gain cross-sections at varying heights within the operating range of a cargo drone \textcolor{black}{are also given according to}  the sectoral antenna pattern as defined in Recommendation ITU-R F.1336-5, \textit{recommends} 3.2.1, \textcolor{black}{for}  the directional antenna gain for peak side-lobe patterns in the frequency range from 6 GHz to 70 GHz. Symmetric elevation and azimuth beamwidths are considered. Even when the beam is pointing almost 10 km away from the depicted region, almost uniform gain are observed \textcolor{black}{towards the ground}. \textcolor{black}{Additionally}, as the likelihood of the presence of a line-of-sight path is high, the performance degrading impact of the small-scale fading is relatively low \textcolor{black}{when} compared to that of  terrestrial networks. A Ricean fading model has often been considered in HAPS connections \cite{Survey}. \textcolor{black}{These} geometric advantages can compensate for the high path loss of the ground-HAPS connections as the HAPS nodes are located at 20 km above ground level. HAPS nodes \textcolor{black}{have} a wide line-of-sight area that rather advantageous while sensing the ground \textcolor{black}{considering location sensing and sensed data collection and gathering}. Additionally, due to the large form factor of HAPS systems that can be equipped with large payloads of high computation capabilities, ML can also play a role in these set-ups, targeting the corresponding optimization problems in terms of radio resource managements and beamforming \textcolor{black}{Additionally, the decisions that need to be made at each fleet element, such as the route and velocity determination can be also done collectively with the data that has been collected at the HAPS node}.

In addition to  coverage, the size of the HAPS  provides the opportunity to allocate computational and caching resources as the payload. These resources will enable quick responses to the connected devices by enabling computing at a close proximity to the fleet elements (as opposed to the cloud data center), which can reduce the overall end-to-end delay \textcolor{black}{specifically with the help of edge computing \cite{e1,e2},} and provide a unified and seamless computation resource for \textcolor{black}{edge/fog} computing.  From a caching perspective, instantaneous traffic data, maps, and  navigation related information can be quickly accessed from the data storage elements in the HAPS node. \textcolor{black}{Sensing under these conditions is also possible through localization and collection of the sensed measurements, including the context related data such as the weather conditions, that may have a significant effect on the performance of the operation of the fleet elements.} Due to the position of a HAPS node in the sky, an unobstructed line-of-sight channel is likely to be encountered. This scenario can provide the full benefit of  massive multi-input multi-output (MaMIMO) architectures. %The building obstructions to be encountered with very limited probability, 
Due to these facts, the use of HAPS to support aerial delivery networks is a natural fit. 

\section{AI-Powered and Connected Aerial Delivery Networks} 

The delivery operation for each delivery fleet member, including cargo drones, is configured independently from the rest of the delivery requirements. To be able to address the ever-increasing demands from consumers,  the scalability of the delivery network needs to be addressed jointly with the corresponding constraints of the fleet elements with a focus on densely populated urban areas. A prominent solution is the aerial delivery networks. A vision of a next-generation delivery network is depicted in Fig. \ref{fig-ML}. The autonomous delivery fleet can be composed of  terrestrial vehicles, marine vehicles, and  aerial vehicles of varying levels of \textcolor{black}{automation} \cite{autonomous}. \textcolor{black}{A} fully autonomous fleet without any human operator is envisioned as the final goal. 

The autonomous delivery fleet operates between the warehouse and the customers \cite{MacKenzie}. The management of this fleet requires ubiquitous connectivity at all times. However, the ubiquitous connectivity alone does not guarantee optimal operations. The advances in AI needs to be exploited for real time operational planning, which may include instantaneous route changes due to impulsive effects, such as weather conditions. \textcolor{black}{A HAPS node can connect the status of the warehouse with the instantaneous conditions of the delivery fleet and the consumer in a low latency setting with the goal of providing a fully automated architecture.  AI \textcolor{black}{can help determine the instantaneous actions that need be taken by all fleet elements, as the environmental conditions and the wireless channel characteristics are expected to be dynamic}.} To enable  low-latency transport network responses in a distributed sense within the autonomous delivery fleet, caching\textcolor{black}{,} computation\textcolor{black}{, and sensing} services need to be accessible by each transportation element\textcolor{black}{.}

\subsection{The Expected Role of Artificial Intelligence (AI)}
The retail sector currently makes active use of AI. For instance, recommendation engines on the consumer side, stocking on the warehouse side,  and side route planning on the delivery  solutions have been well-studied problems, and customized solutions are already available\textcolor{black}{.} However, the integration of a real-time  autonomous fleet and the full 3D connectivity along with caching\textcolor{black}{, sensing and computation} offloading functionalities can enable a multi-sector AI implementation that \textcolor{black}{can}  also available to low-latency responses to the changing conditions and  requirements. % A real-time AI across the domains is not yet implemented.  

Coherent operations between a warehouse, aerial delivery network,  ground delivery network and the consumer can be enabled through the use of multi-faceted ML techniques. \textcolor{black}{The operational goals may require conflicting objectives such as jointly minimizing the transmission time and minimizing the total energy both at the fleet element level and the system level.} %As a tradeoff, a weighted combination of the maximum delivery transmission time, a long-term-goal, and the total required energy of the fleet elements can be used as an objective function. 
 \textcolor{black}{The objective function can be determined based on the communication aspects (such as minimizing the maximum delay and the total energy consumption of each fleet element) or from an application level perspective (such as minimizing the overall packet delivery time for a particular customer).} Based on the inherent complexity of the corresponding optimization problem, the ML-based solutions can be a practically feasible alternative to be deployed across different sectors. %A distributed learning architecture is expected to pave the way for this multi-industry perspective. 
In this optimization problem,  the operating states can be jointly processed with the demand forecast even instantaneously, based on  \textcolor{black}{the consumer behaviour that may change instantaneously}.

\textcolor{black}{ML-based solutions require a representative labeled set of data, sufficiently large for training. However, in the HAPS service architecture, the conditions of the wireless communication channels and the air transport environment change frequently. Continuously learning and re-optimizing with supervised learning will require precious computational resources and time for the complete system. Therefore techniques such as active learning (AL) will be beneficial. AL is about AI asking questions to learn. The central AI unit will decide which nodes in the system will share wireless channel measurements and share air transport environment measurements at each time interval. The sampling decision can be based on information-maximization techniques. As a result, the re-optimization and disruption management with AI can be deployed in an energy-efficient and time-efficient manner.} 

Despite the apparent benefits,  the joint considerations of demand and route planning have not yet been implemented due to the associated complexity and  technical limitations. However, 3D connectivity offers the potential to enable each of the delivery nodes to behave in an autonomous manner for  supply chain management and  delivery via computation\textcolor{black}{,} caching\textcolor{black}{, and sensing} services. We envision that the ML approaches will serve to connect the supply management in the warehouse,  delivery network along with   instantaneous \textcolor{black}{consumer} demands. 

As it is expected that a single model may not accurately help solving the problem, ensemble learning techniques can be of benefit \cite{MLSurvey}\textcolor{black}{, especially considering the individual conditions of the fleet elements}. Distributed learning approaches, including the parameter server paradigm, federated learning, or the fully distributed sufficient factor broadcasting approaches, can serve as powerful tools \textcolor{black}{that can help with the decision of the next action at each fleet element, both in terms of their location and navigation information and the transmit power levels} to enable this vision\textcolor{black}{.} \textcolor{black}{The conditions of the wireless channel, either centered at the HAPS node, or through distributed communication via device-to-device (D2D) connectivity of fleet elements need to be closely monitored to capture the dynamic environmental/channel conditions}. Such a network has the potential to increase  customer satisfaction by addressing their needs as fast as possible, along with the OPEX reduction in terms of  delivery architecture. Further savings from the fuel consumption perspective is also possible, \textcolor{black}{as numerically shown in} \cite{Hanzo}. 

\subsection{Operating Aerial Networking Components in Urban Areas}

Autonomous delivery fleets \textcolor{black}{will} include cargo trucks, vans, motors, cargo planes, autonomous or human operated VTOL aircraft, and autonomous drones for parcel delivery. The ranges and  energy efficiencies of these devices are determined according to the properties of individual fleet element. For example, autonomous ships mainly operate over  designated seaways. Trucks mainly operate on highways. The optimization of the route planning for heterogenous fleet components is currently  under investigation and promising results have been already recorded \cite{Optimizasyon}. \textcolor{black}{Most}  delivery network elements have predetermined operational characteristics, whose boundaries are still under development for  aerial delivery network, including VTOL aircraft and drones. The aerial delivery network needs to be coherently operational with the ground network, which is also composed of conventional delivery trucks with drivers and autonomous vehicles of varying sizes.

\begin{figure}[tb!]
    \centering
    \includegraphics[width=0.98\linewidth]{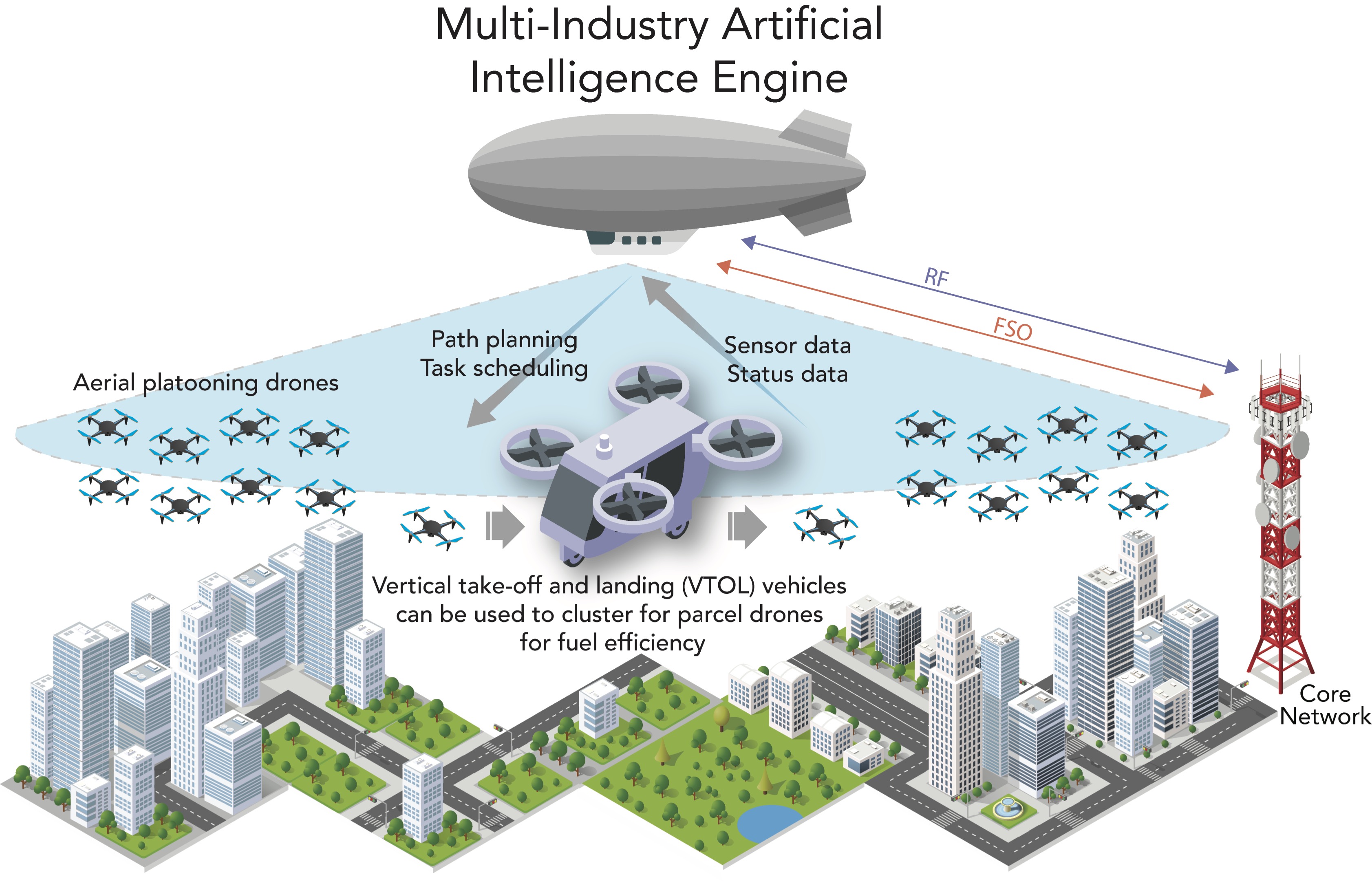}
\caption{A hierarchical aerial platooning is depicted, where  coverage, computation, caching, and sensing services are provided by a HAPS.} \label{Platoon}  
\end{figure}

\subsubsection{Aerial Platooning}

Platooning in vehicular networks aims to control multiple vehicles on the basis of a leading vehicle and the use of cruise control. The following vehicles adjust their speeds and paths on the basis of the leading vehicle. The fuel savings and  increased traffic efficiency benefits make platooning an attractive paradigm in today’s intelligent transportation systems. Next generation delivery networks can also benefit from the advantages provided by platooning in 3D aerial settings. The leading aerial vehicles can be VTOL aircraft or higher capacity drones. From the perspective of battery power and fuel, the path and  task planning can be instantaneously executed in the HAPS node, and the corresponding flight commands can be transferred back to the platoon’s leading aerial vehicle. Also instantaneous traffic data can be cached at the HAPS node, which can enable the generation of  accurate path and task commands. The envisioned \textcolor{black}{aerial platooning} operation is depicted in Fig.~\ref{Platoon}\textcolor{black}{, where} on the basis of available fuel and battery resources of the cargo drones, co-transportation on the higher capacity aerial vehicles, such as VTOL aircraft, can also be planned instantaneously, depending on the final destination of the cargo drones. For the task planning and  route planning, the potential of the reinforcement learning techniques are visible even from the terrestrial vehicular platoons \cite{RL-platoon}. \textcolor{black}{These capabilities will enhance the multi-industry artificial intelligence-based management of the next generation aerial delivery network, through improved energy efficiency and sustainability. } 

\begin{figure}[tb!]
    \centering
    \includegraphics[width=0.98\linewidth]{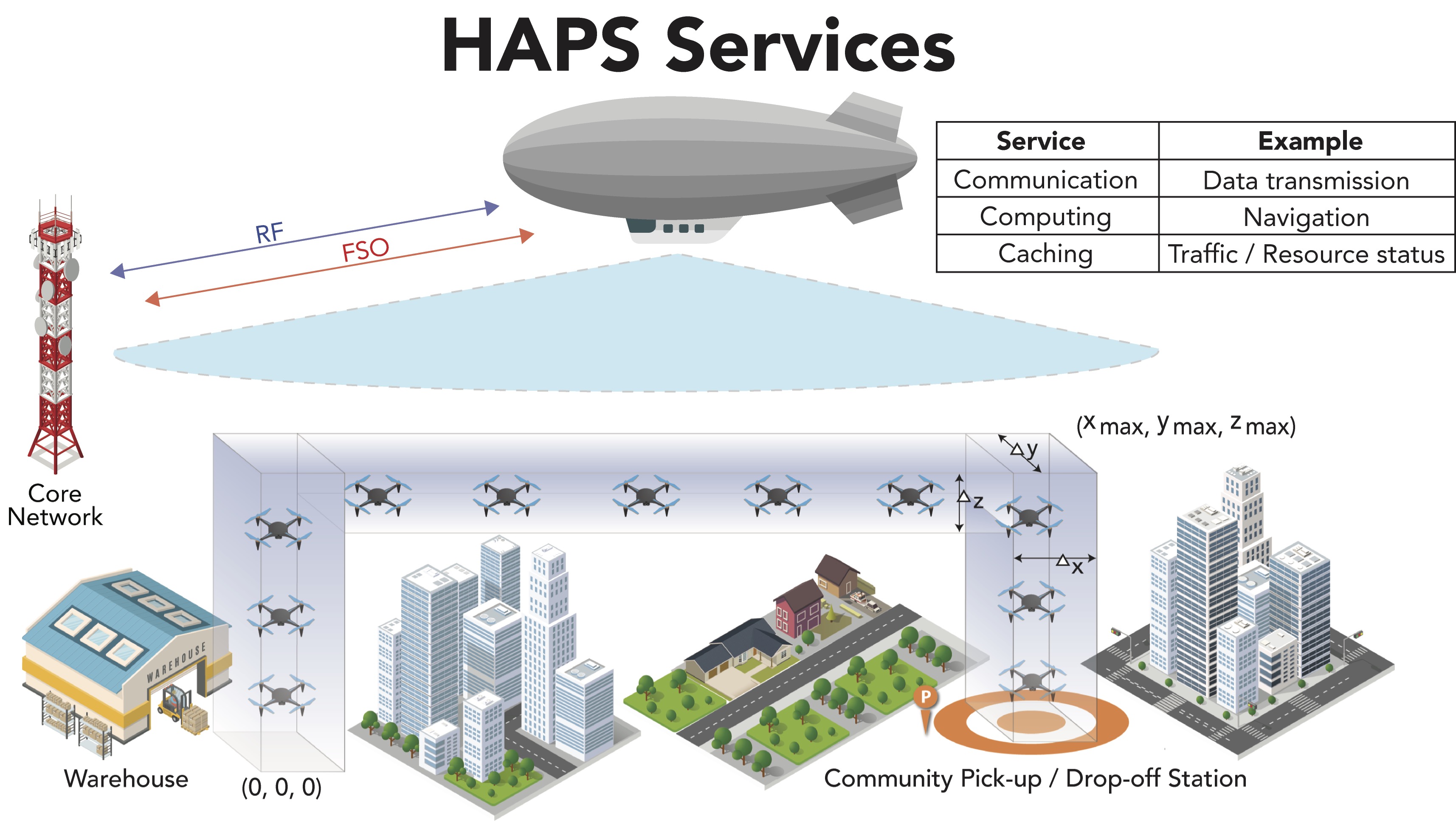}
    \caption{The 3D aerial highway from the warehouse to a metropolitan area is depicted. The community pick-up/drop-off stations enable a relatively simpler path planning for the final part of the route. The connectivity of the aerial devices is maintained by the HAPS.} \label{community}
\end{figure}

\subsubsection{Community Pick-up/Drop-off  Stations}
A challenge for  drone delivery nodes will be reaching consumers at their homes. Unlike traditional delivery services, which can deliver a package to a consumer with the ring of a doorbell, drone delivery nodes may face challenging operational environments. For these cases, we envision community pick-up/drop-off points where  packages can be picked up, and notifications for the arrival of packages can be sent in advance, in accordance with user preferences,  for example 15 minutes before the expected delivery time. \textcolor{black}{This application is described in Fig. \ref{community}, along with the possible communication, computing,   caching, and sensing services that can be provided by the serving HAPS node including. During their flight times over the aerial highways, the drones can capture instantaneous changes in the navigation and/or  routing using the information either processed  or stored by the HAPS node.}  These community \textcolor{black}{pick-up/drop-off  stations} can also be used to transfer other packages that may be picked up by the autonomous drones or droids. 

\subsubsection{3D Aerial Highways}
It is expected that the number of aerial delivery vehicles will  increase. For the sustainable management of these vehicles, along with  corresponding regulations, proper route planning is a must. To help in planning, 3D highways are being considered by  NASA  and FAA under the UTM activities \cite{UTM}. The guidelines and regulations aim to provide a monitored speed region  for metropolitan deliveries. This solution evokes the famous 1980s cartoon series, the Jetsons.
 
\begin{figure}[t]
    \centering
    \includegraphics[width=1\linewidth]{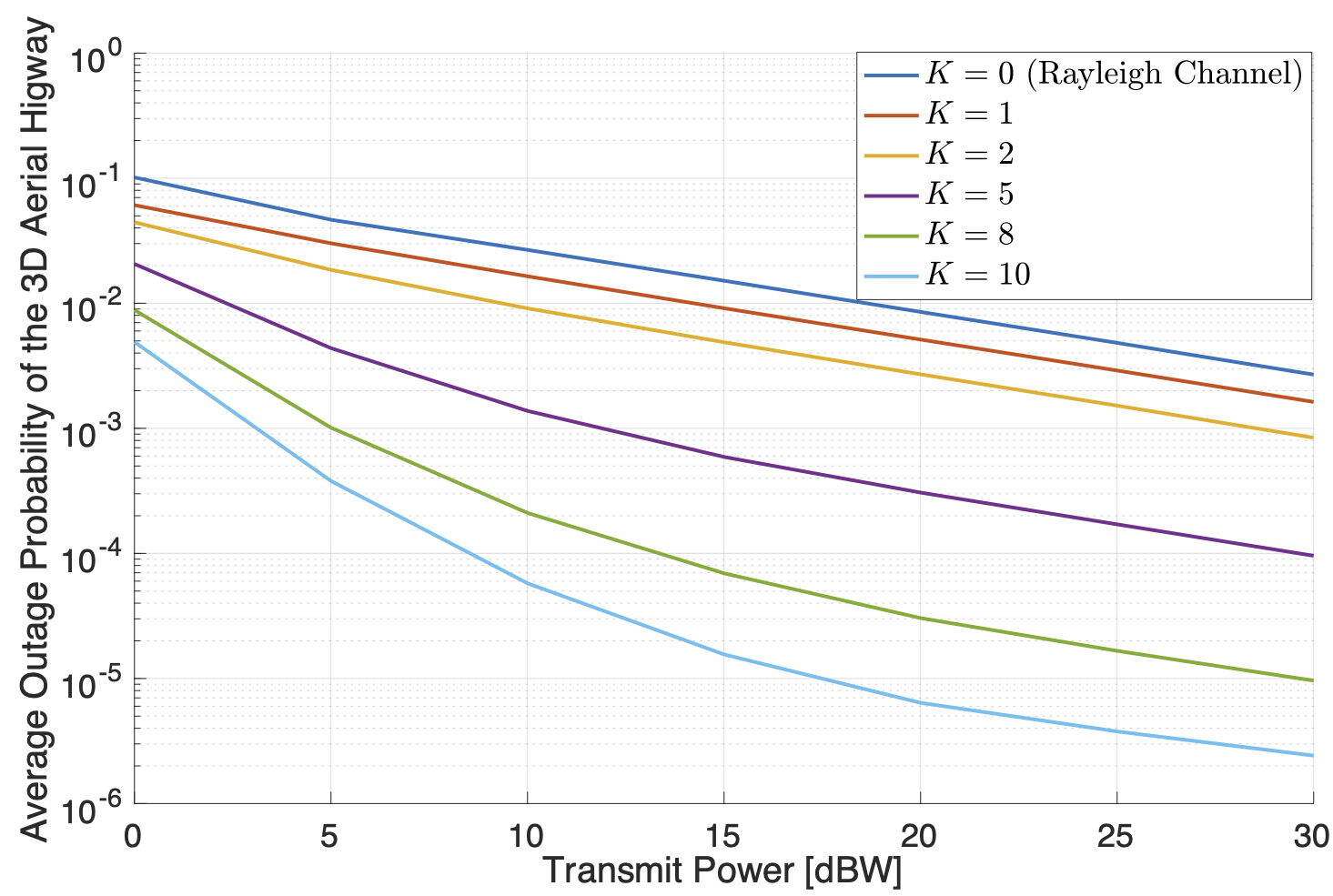}
\caption{The volumetric average outage probability of the 3D aerial highway. %As the line-of-sight component (K parameter) increases, the performance changes favorably as expected from the models of the HAPS-to-ground channel models. 
}\label{fig-HAPS2}
\end{figure}

Mimicking the highway/street hierarchy of terrestrial roads, multiple regulated speed limits within these drone highways can make the operation of multiple fleet elements possible, which can provide scalable retail solutions for  next generation consumer networks. The quasi-structured mobility restrictions will facilitate the operations of \textcolor{black}{a} drone fleet \textcolor{black}{with mobile fleet elements}, especially in densely populated areas \textcolor{black}{due to the large coverage area that may be provided by the HAPS node thanks to its high altitude}.  The ubiquitous access to the optimization engine supported by AI will enable near real-time planning over these 3D highways, enabling the instantaneous reactions of the drones while operating.  The availability of the navigation and the route planning data will also provide an increased level of safety and reliability. The advantageous channel characteristics of the 3D coverage provided by a HAPS are clearly shown in Fig. \ref{fig-HAPS2}.  In line with the models in the literature, varying line-of-sight power (K parameter) values are considered for a Ricean channel, and the average outage probability values are shown for an aerial highway of the dimensions noted in Fig. \ref{community}. The corresponding simulation parameters are given in Table \ref{sim}. \textcolor{black}{The average outage probabilities of the volumes of the 3D aerial highway\footnote{\textcolor{black}{Dimensions of this 3D aerial highway are given in Fig. 4, according to the parameters given in Table I.}},  plotted in Fig. \ref{fig-HAPS2}, according to the changing channel conditions show that, as the K value increases an outage becomes less excepted. Yet even in the absence of a line of sight (K$=0$), the average outage can be adjusted according to the desired level by changing the transmit power value. } The terrestrial base stations simply cannot provide coverage at this height, as opposed to the acceptably low outage probabilities that a HAPS can provide in a 3D aerial highway.


%\IEEEPARstart{T}{his} demo file is intended to serve as a ``starter file''
%for IEEE journal papers produced under \LaTeX\ using
%IEEEtran.cls version 1.8b and later.

%I wish you the best of success.

 

\begin{table}
\caption{Simulation parameters.}
\label{sim}
\begin{tabular}{|p{3cm}|p{5cm}| }
\cline{1-2}
\textbf{Parameter} & \textbf{Value}    \\ \cline{1-2}
Carrier frequency & 10 GHz    \\ \cline{1-2}
Bandwidth &  10 MHz\\ \cline{1-2}
Temperature & 24 $^\circ$C \\ \cline{1-2}
Normalized Rate & 1 b/s/Hz \\ \cline{1-2}
Channel  & Ricean \textcolor{black}{K$\in \{0,1,2,5,8,10\}$}\\ \cline{1-2}
HAPS Antenna Pattern & ITU-R F.1336-5, \textit{recommends} 3.2.1\\ \cline{1-2}
\textcolor{black}{HAPS Height} & \textcolor{black}{20 km}\\ \cline{1-2}
HAPS Beam & Pointing towards  10Km in $x$ direction 500 m in $y$ direction\\ \cline{1-2}
Dimensions of the 3D Aerial Highway  & $\Delta x = 10$ m, $\Delta y = 10$ m,  $\Delta z = 10$ m, $X_{\textrm{m}} = 100$ m, $Y_{\textrm{m}} = 10$ m,  $Z_{\textrm{m}}= 100$ m \\ \cline{1-2}
 \end{tabular}
\end{table}

\section{Open Issues}

The relevant open research problems are discussed below. 
\paragraph*{Seamless integration of the delivery network and the VHetNet} Overall operations of the integration of the next generation delivery networks and the next generation ICT network, VHetNet, need to be seamless from the end-user perspective \textcolor{black}{according to the fleet elements}. This level of seamless integration needs additional care\textcolor{black}{ in terms of the connectivity and the energy levels of the fleet elements} rather than simply assigning a network slice corresponding to the delivery services. \textcolor{black}{ For instance, for the fleet elements with insufficient energy levels, platooning options  that may serve energy may be initiated. }  The related services can be offered via a multi-industry standardization activity among retailers and communications service providers. Furthermore, security, privacy, and safety of the flights also need to be monitored by the same entity. 

\paragraph*{Joint connectivity between the fleet, edge, and core elements} Functionalities of the aerial fleet elements need to be identified for network agility aspects. The waveform designs and the use of  non-contiguous bandwidth is an interesting challenge that has not been encountered before. The clustered aerial platooning architecture forces the use of device-to-device links, whereas the edge connection at the HAPS and  core-based cloud computing functionalities also need to be supported. Due to the geometric advantages provided by the HAPS nodes \textcolor{black}{based on their high amplitudes}, extremely narrow \textcolor{black}{3D} pencil beams can be used for high data rate connections to the fleet elements\textcolor{black}{, while aiding the related interference management issues}. Inter-HAPS handoff strategies also need to be investigated for a worldwide deployment. \textcolor{black}{The mobility management for the high number of fleet elements also needs to be addressed for a successful deployment.}

\paragraph*{Cognitive radio resource management} Enhanced cognition capabilities that are dependent  not only on the  spectrum usage status, but that also include the energy storage aspects of the individual components of the fleet elements are needed. This will eliminate a high control plane load on the serving HAPS node. As a single HAPS node will provide service to tens of kilometers, even high-speed aerial nodes can be served with a single HAPS cell without the need for a sophisticated mobility management framework, so that these high-speed  nodes can remain operational in these selected frequency bands after performing spectrum sensing. The use of higher frequency bands, including the terahertz bands, can be a remedy to alleviate potential packet collisions\textcolor{black}{.}% in the cognitive interfaces.  

\paragraph*{Computation algorithms at the edge} The computation algorithms to address instantaneous customer demand along with the instantaneous changes in the fleet management\textcolor{black}{. They} also perform fleet management via path planning and task scheduling while addressing the connectivity in the VHetNet. Customized algorithms to address this multi-industry operation are needed to enable a scalable extension of the targeted services in metropolitan areas. 

\paragraph*{Energy management} There are two perspectives to energy management in next generation delivery networks. From the communication perspective, the HAPS nodes are always considered a green solution, for they mainly extract energy from solar panels \cite{Survey}. They also use hybrid energy sources, including wind and solar energy, along with possible RF energy harvesting approaches. Maintaining a quasi-stationary position against strong winds and varying weather conditions requires supplementary energy in addition to the energy needed for the payload. \textcolor{black}{Improved energy efficiency levels  will be needed, even} %Hence, improved efficiency levels  will be needed in terms of both harvesting and storage. Even 
nuclear energy may  be an option for a sustainable operation. 

%Considering the delivery network perspective, the energy management of the drones will be continuously monitored. In case of  emergency power needs, simultaneous wireless information and power transfer (SWIPT) based energy transfer from the HAPS node to the fleet elements may be an effective approach. The potential savings in the delivery network from OPEX and CAPEX perspectives need to be quantified in a realistic manner. %\newpage 

\section{Conclusions}
A new wireless network architecture is needed to enable the functionality of a fully autonomous parcel delivery network with aerial fleet elements, including  cargo VTOL aircraft and  cargo drones. In this paper, we presented our vision of a network that not only assists with sensing capabilities while providing  reliable connectivity, but also serves as a computational and caching platform, powered by the HAPS systems of the VHetNets. 

%\balance

\section*{Acknowledgment}

The authors would like to thank Prof. Abbas Yongacoglu for the valuable discussions.  This work was supported by Huawei Canada Co., Ltd.
\bibliographystyle{IEEEtran}
\bibliography{references}

%\end{thebibliography}


%\end{IEEEbiography}


%\newpage



%\vfill



\section*{Biographies}
\footnotesize{
%\vspace{-0.1cm}
GUNES KARABULUT KURT [StM'00, M'06, SM'15] (\textcolor{black}{gunes.kurt@polymtl.ca}) received the Ph.D. degree in electrical engineering from the University of Ottawa, Ottawa, ON, Canada, in 2006. Between 2005 and 2008, she was with TenXc Wireless, and Edgewater Computer Systems, in Ottawa Canada. From 2008 to 2010, she was with Turkcell R\&D Applied Research and Technology, Istanbul. \textcolor{black}{G. Karabulut Kurt is with the Department of Electrical Engineering,  Polytechnique Montr\'eal, Montr\'eal, Canada}. She is also an Adjunct Research Professor at Carleton University. She is serving as an Associate Technical Editor of IEEE Communications Magazine.\\

%\vspace{-0.1cm}
HALIM YANIKOMEROGLU [F] (halim@sce.carleton.ca) is a full professor in the Department of Systems and Computer Engineering at Carleton University, Ottawa, Canada. His research interests cover many aspects of 5G/5G+ wireless networks. His collaborative research with industry has resulted in \textcolor{black}{37} granted patents. He is a Fellow of the Engineering Institute of Canada and the Canadian Academy of Engineering, and he is a Distinguished Speaker for IEEE Communications Society and IEEE Vehicular Technology Society.}

%GUNES KARABULUT KURT [StM'00, M'06, SM'15] (gkurt@itu.edu.tr) received the Ph.D. degree in electrical engineering from the University of Ottawa, Ottawa, ON, Canada, in 2006. Between 2005 and 2008, she was with TenXc Wireless, and Edgewater Computer Systems, in Ottawa Canada. From 2008 to 2010, she was with Turkcell R\&D Applied Research and Technology, Istanbul. Since 2010, she has been with ITU. She is also an Adjunct Research Professor at Carleton University. She is serving as an Associate Technical Editor of IEEE Communications Magazine.\\

%GUNES KARABULUT KURT [StM'00, M'06, SM'15] (gkurt@itu.edu.tr) received the Ph.D. degree in electrical engineering from the University of Ottawa, Ottawa, ON, Canada, in 2006. Between 2005 and 2008, she was with TenXc Wireless, and Edgewater Computer Systems, in Ottawa Canada. From 2008 to 2010, she was with Turkcell R\&D Applied Research and Technology, Istanbul. Since 2010, she has been with ITU. She is also an Adjunct Research Professor at Carleton University. She is serving as an Associate Technical Editor of IEEE Communications Magazine.\\
\end{document}